\documentclass{aa}

\usepackage{lineno}
\usepackage{hyperref}
\usepackage{siunitx}
\usepackage{booktabs}
\usepackage{tabularx}
\usepackage{array}
\usepackage{graphicx}
\usepackage{subfig}
\usepackage{afterpage}
\usepackage{amsmath}
\usepackage{upgreek}

\usepackage{txfonts}
\usepackage{hyperref}

\begin{document}

   \title{Great comet C/2023 A3 (Tsuchinshan–ATLAS): dust loss before perihelion}

   \author{Bin Liu
          \inst{1}
          \and
          Man-To Hui\inst{1}
          \and
          Xiaodong Liu\inst{2}
          }
   \institute{State Key Laboratory of Lunar and Planetary Sciences, Macau University of Science and Technology, Taipa, Macau 999078, People's Republic of China \and School of Aeronautics and Astronautics, Shenzhen Campus of Sun Yat-sen University, Shenzhen, Guangdong 518107, People's Republic of China  \\
            \email{{mthui@must.edu.mo,
            manto@hawaii.edu}
             % \thanks{}
             }}

   \date{}

  \abstract
{In this study, the dust loss of comet C/2023 A3 (Tsuchinshan–ATLAS) is investigated through the analysis of archival images. By measuring the surface brightness profile of the coma, we determined that the comet maintained nearly in a steady state during the observations. Analysis of the dust distribution perpendicular to the orbital plane reveals that the ejection velocity is $v_{\perp}\sim(65\pm5)\,\beta^{1/2}$ m s$^{-1}$, where $\beta$ is inversely proportional to the size of the dust grains. From the dust scattering cross-section measurement, we estimated the upper limit of the nucleus radius to be $\sim\!5.9\pm0.2$ km, assuming a geometric albedo of 0.04. Based on the extrapolation of the scattering cross-section over time, the onset time of significant dust activity is estimated to be 25 July 2022, corresponding to a heliocentric distance of 9.1 au, with the activity mechanism at this distance likely being the phase transition from amorphous to crystalline ice. Our simulation reveals that the minimum dust size is \SI{20}{\micro\meter} and the size distribution index is $s = 3.4$ in tail. The dust loss rate is determined to be $(1.7 \pm 0.8) \times 10^2$ kg s$^{-1}$, based on the derived average size of the particles and the rate of change of the scattering cross-section over time. Through a simplistic model, we evaluate that the nucleus of the comet remains stable against tidal effects, sublimation, and rotational instability, and disfavour the fate of disintegration. The result is consistent with observations that the nucleus has survived.}


   \maketitle
%
%-------------------------------------------------------------------

\section{Introduction}
Comets as primordial bodies of the solar system contain material clues from its early stages of formation \citep{betzler2012nonextensive}. They offer a vital window into studying the origin and evolution of the solar system \citep{bailey1996provenance}. Long-period comets originate from the distant Oort Cloud \citep{dones2004oort}. Their extended orbital periods and limited external disturbances allow them to reflect the early solar system's environment more accurately \citep{hands2020capture}. Recent advances in astronomical observing techniques have enabled the discovery and detailed study of many new comets. Among them, C/2023 A3 (Tsuchinshan–ATLAS) stands out for its unique orbital characteristics and striking brightness. 

The long-period comet C/2023 A3 (Tsuchinshan–ATLAS), hereafter "A3", was discovered on 9 January 2023 by the Purple Mountain Observatory in China at a magnitude of 18.7, and independently identified on 22 February 2023 by the Asteroid Terrestrial-impact Last Alert System (ATLAS) in South Africa at a magnitude of 18.1 \citep{ye2023comet}. The orbit of A3 is characterized by a perihelion distance of 0.391 au, an eccentricity of 1.00014, an inclination of 139.1°, a longitude of ascending node of 317.3°, and an argument of perihelion of 59.0° (based on Minor Planet Center data as of June 2024).

A3 displayed a bright cometary tail and a thin anti-tail and has since been classified as the Great Comet of 2024 \citep{moreno2025cometary}. \citet{sekanina2024inevitable} suggested a potential disintegration event might have occurred to the comet based on an anomalous peak in its light curve on 15 April 2024 at 3 au followed by a systematic decline in dust production. \citet{grant2024prospects} predicted that Earth would cross its ion tail between October 10 and 13, 2024, with a minimum distance of $(1.47\sim3.89)\times 10^6$ km, an event that could potentially be detectable by spacecraft at the Sun-Earth L1 point and might influence the terrestrial magnetosphere \citep{grant2024prospects}. Furthermore, a carbon-depleted composition has been indicated by spectral analyses, with only sodium D-lines detected and no carbon-bearing species such as $\text{C}_2$ or $\text{C}_3$ observed, corroborated by molecular production rates showing a low $\text{C}_2$/CN ratio \citep{tang2024spectrum,ahuja2024molecular,jehin2024trappist}. The analysis of dust loss from A3, given its unique characteristics, not only helps in evaluating the possibility of comet fragmentation and identifying the mechanisms that trigger such events but also yields valuable data on dust activity in carbon-depleted comets. This contributes to a clearer picture of how cometary material is released. Furthermore, these findings can guide the development of dust detection instruments for future missions, improving their capacity to detect and analyse cometary dust.
\begin{figure*}
\centering
    \includegraphics[width=2\columnwidth]{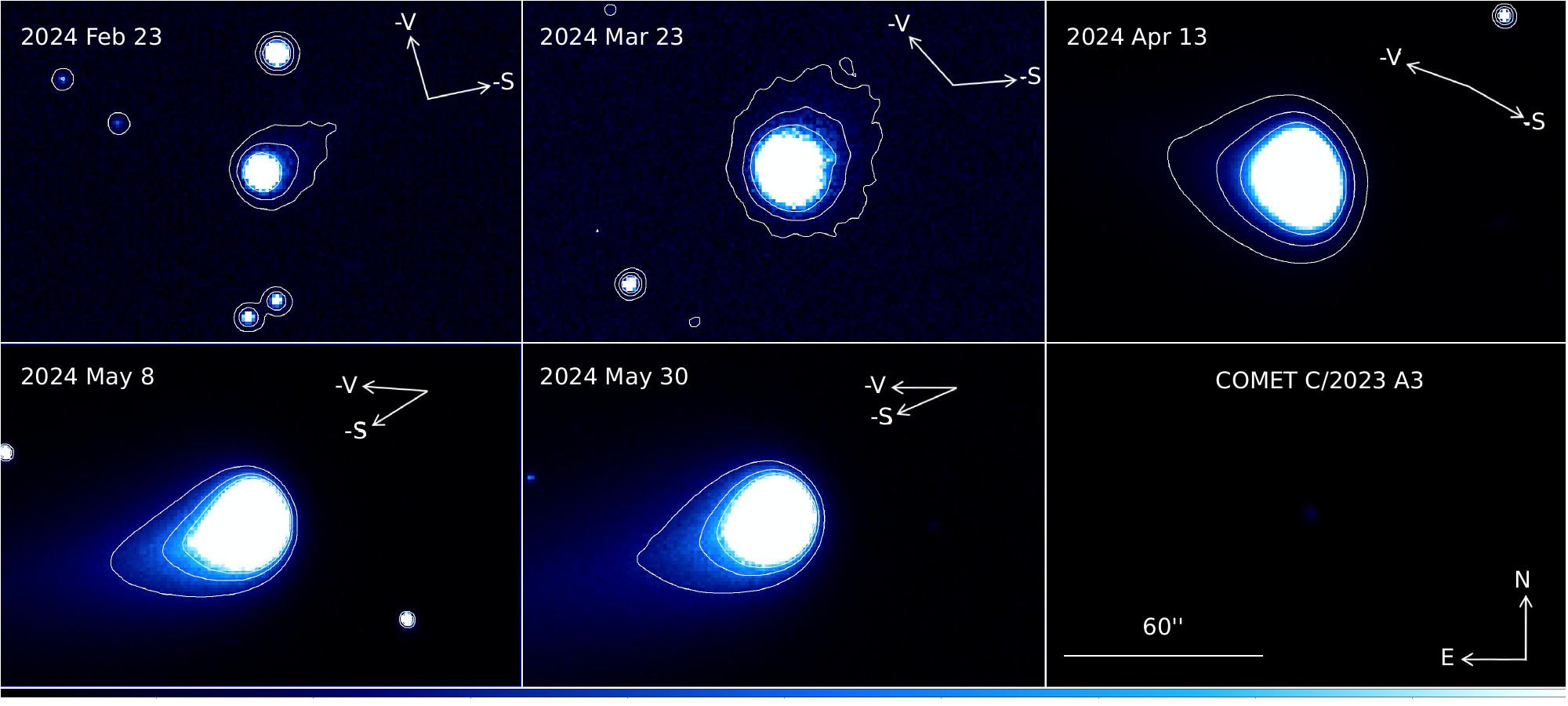}
    \caption{Composite ZTF images of A3, with the upper right corner of each panel indicating the observation epochs. White arrows denote the projected Sun-A3 direction (-S) and the opposite of A3’s heliocentric velocity (-V).}
    \label{ObsImag}
\end{figure*}

In this study, we analyse archival observation images of A3 and employ a dust dynamics model to reproduce its dust morphology, enabling characterization of its dust loss. The paper is structured as follows. The archival observation images and the data reduction process are presented in Section \ref{Data}. The results of morphological and photometric analyses are summarized in Section \ref{Results}. The dust properties and the stability of the nucleus are discussed in Section \ref{discussion}. Finally, the conclusions of this paper are listed in Section \ref{CONCLUSIONS}.

\section{Data}
\label{Data}
Archival images of A3 was retrieved from the Zwicky Transient Facility (ZTF) at Palomar Observatory. ZTF is designed to detect transient objects such as supernovae, asteroids, and comets. It employs a 48-inch Schmidt telescope fitted with a 600-megapixel CCD camera, providing a 47 deg$^2$ field of view and an angular resolution of 1\arcsec per pixel in each exposure \citep{graham2019zwicky}.

Figure \ref{ObsImag} presents five representative r‑band frames—each a 30‑second exposure—captured on 23 February, 23 March, 13 April, 8 May, and 30 May of 2024, during which A3 exhibited ongoing activity. For uniformity, all images are cropped to a $2.5\arcmin\times2.5\arcmin$ field. Data processing is performed using SAOImage DS9 \citep{joye2003new} and AstroArt \citep{nicolini2003astroart}. Table~\ref{geometry} lists the observing geometry for each epoch.

\begin{table*}
    \centering
    \renewcommand{\arraystretch}{1.15}
    \caption{Observing geometries of A3}
    \label{geometry}
    \setlength{\tabcolsep}{12pt}
    \begin{tabular}{lcccccccc}
		\toprule
Date & DOY$^a$ & $R^b$ & $\Delta^{c}$ & $\alpha^d$ & PsAng$^e$ & PsAMV$^f$ & $\nu^{g}$ & $\delta^{h}$ \\
      (UT)  &(days) & (au) & (au) & (deg) & (deg) & (deg) & (deg) & (deg)  \\
		\hline
23 Feb 2024 & 54 & 3.63 & 3.23 & 15.1 & 283.0 & 17.1 & 218.3 & 8.6 \\
23 Mar 2024 & 83 & 3.27 & 2.44 & 11.1 & 275.6 & 43.5 &  220.5 & 5.0 \\
13 Apr 2024 & 104 & 3.00 & 2.02 & 3.9 & 241.9 & 71.5 & 222.3 & -0.5 \\
8 May 2024 & 129 &  2.67 & 1.77 & 12.1 & 122.3 & 87.2 &  225.0 & -9.4 \\
30 May 2024 & 151 & 2.36 & 1.79 & 23.5 & 114.6 &  91.0 &  228.1 & 15.8 \\
		\hline
		\multicolumn{9}{l}{$^a$ Day of the year, with 1 January 2024, as day 1.}\\
		\multicolumn{9}{l}{$^b$ Heliocentric distance of A3.}\\
		\multicolumn{9}{l}{$^c$ Geocentric distance of A3.}\\
		\multicolumn{9}{l}{$^d$ Phase angle, Sun-A3-Earth.}\\
		\multicolumn{9}{l}{$^e$ Position angle of the Sun-A3 direction.}\\
		\multicolumn{9}{l}{$^f$ Position angle of the vector opposite to the projected heliocentric velocity.}\\
		\multicolumn{9}{l}{$^g$ True anomaly of A3.}\\
		\multicolumn{9}{l}{$^h$ Angle between the observer’s viewing direction and A3’s orbital plane.}\\
	 \end{tabular}
    \end{table*}

\section{Results}
\label{Results}
\subsection{Morphology}
\label{morphology}
Figure \ref{ObsImag} shows that A3 is surrounded by a relatively symmetrical inner coma with a radius of approximately 5\arcsec, which presumably reflects isotropic dust ejection of the nucleus. As the coma expands outward, particles in the coma are pushed in the anti-solar direction by radiation pressure, supplying material to the dust tail. The observation from 23 February 2024 shows that the tail fades to the sky background at $\sim\!30\arcsec$ from the nucleus, corresponding to a projected distance of about $7\times10^4$ km. During the observed period, the dust tail gradually lengthens. As of the 30 May 2024 observation, the dust tail extends far beyond the image field of view, which is 60\arcsec~(corresponding to a projected distance of approximately 80000 km), indicating that the tail's length exceeds this visible limit.

To characterize the surface brightness distribution of the coma near the nucleus, we focus on the inner region (within 30\arcsec), which allows for detection of most of the coma's signal while excluding the fainter dust tail. The surface brightness profile of the coma, $\sum(\theta)$, expressed as a function of angular radius, $\theta$, is determined using 30 concentric annular apertures-each 1\arcsec wide-centered on the coma’s optocenter. The background brightness is determined using an annular aperture, spanning from 80\arcsec to 120\arcsec in radius. The surface brightness profiles are first measured from the images taken on May 7 and 30, as they provide the best signal-to-noise ratios in the dataset. In most directions around the nucleus, the coma's dominant brightness is concentrated within roughly 30\arcsec from the nucleus, while the tail extends beyond 60\arcsec. By taking the median brightness in each annular aperture, the contribution from the extended tail is effectively excluded, so that the measured brightness primarily reflects the coma’s intrinsic profile.
%02252304

\begin{figure}
\centering
        \includegraphics[width=0.85\columnwidth]{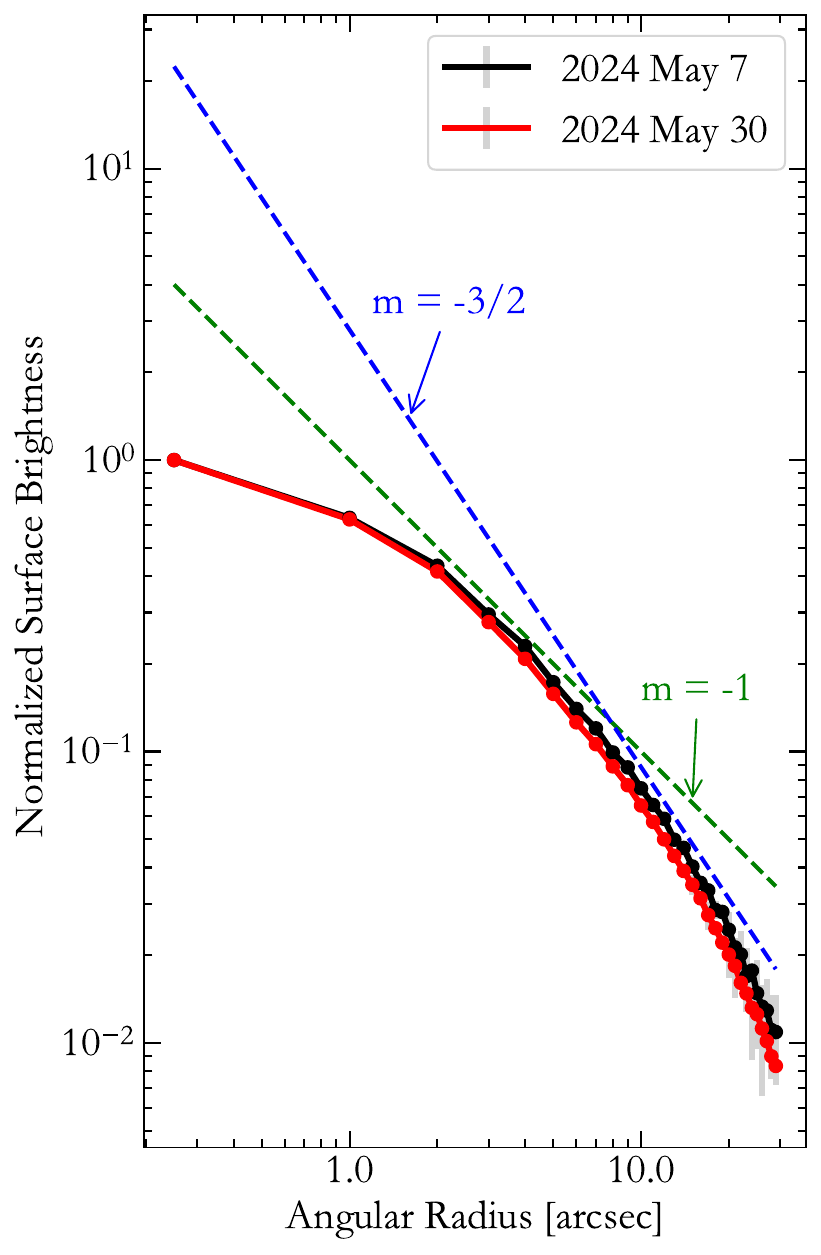}
    \caption{Coma brightness profiles on 7 May 2024 (black curve) and 30 May 2024 (red curve). The blue and green dashed lines denote logarithmic gradients with slopes of $q = -3/2$ and $q = -1$, respectively.}
    \label{ComaSur}
\end{figure}

The measured brightness profiles (black symbols for 7 May and red symbols for 30 May) are shown in Figure~\ref{ComaSur}, from which we can notice that the profile of A3's coma exhibited little to no significant change during May 2024. The brightness profiles of the coma in Figure~\ref{ComaSur} are characterized by three distinct components. In the $0<\theta<2\arcsec$ region, the observed profile is influenced by the convolution of the cometary structure with the point-spread function (PSF) of the ZTF, which has a full width at half maximum (FWHM) of 2.6\arcsec for the May 7 observation and 2.4\arcsec for the May 30 observation. In the $2\arcsec<\theta<7\arcsec$ region, the surface brightness distribution exhibits a power-law behavior, expressed as $\sum(\theta)\propto \theta^{q}$. Our analysis yields a power-law index of $q = -1.02\pm 0.02$ on 7 May 2024 and $q = -1.04\pm 0.03$ on 30 May 2024. These results are statistically consistent with each other. The mean value of $q = -1.03\pm 0.03$ is also consistent with $q = -1$, which is expected for a steady-state coma without solar radiation pressure. In the region where $\theta > 7\arcsec$, as the coma expands outward, radiation pressure gradually becomes the dominant force acting on the dust particles, pushing them into the tail. This results in the progressive steepening of the surface brightness distribution, with the slope $q$ increasing over radial distance until it reaches a value of $q = -1.5$.

For A3, the measured surface brightness profiles demonstrate remarkable agreement with expectations. Specifically, the inner coma has slopes $q$ approximately equal to -1, aligning with the predictions for a steady-state coma without radiation pressure. Meanwhile, the outer region exhibits slopes $q$ around -1.5, as anticipated for regions dominated by solar radiation pressure, and no significant slopes steeper than -1.5 are observed. This suggests that there are no major changes in the dust properties, as deviations steeper than -1.5 are typically caused by the presence of nonsteady-state emission or grain fading. The radial distance of $7.0\pm0.5\arcsec$ at which the inner and outer regions intersect is determined by the fits for $q=-1.5$ and $q=-1$, corresponding to the extent of dust coma in the sunward direction. Following \cite{jewitt1987surface} and \citet{kim2020coma}, this distance $l_{\text{coma}}$, at which solar radiation pressure decelerates dust particles ejected from the comet with initial velocity $v_{\mathrm{ej}}$ to near-zero velocity at the sunward edge of the coma, is given by
\begin{equation}
l_{\text {coma }}=\frac{v_{\mathrm{ej}}^2}{2 \beta g_{\text{sun}}}.
	\label{vej}
\end{equation}
The local gravitational acceleration caused by the Sun is expressed as $g_{\text{sun}} = g {(1)}\sin \alpha/R^2$, where $g{(1)}=0.006$ m/s$^2$ denotes the Sun's gravitational acceleration at 1 au from the Sun, $\alpha$ is the phase angle, and $R$ refers to the heliocentric distance in astronomical units. Additionally, radiation pressure reduces the net force on a dust grain by a factor $\beta$, which scales approximately as the inverse of the particle’s radius. In the observation on 30 May 2024, the turnaround distance, which indicates the sunward extent of the coma, is measured as $l_{\text {coma }}\sim(9.1 \pm 0.6) \times 10^3$ km. The phase angle $\alpha$ and heliocentric distance $R$ on 30 May 2024 can be seen in Table \ref{geometry}. Substituting these parameters into Equation \ref{vej}, the dust ejection velocity $v_{\mathrm{ej}}$ is derived as a function of $\beta$ by $v_{\mathrm{ej}}\sim(88\pm3)\,\beta^{1/2}$ m s$^{-1}$.

\begin{figure}
\centering
        \includegraphics[width=1\columnwidth]{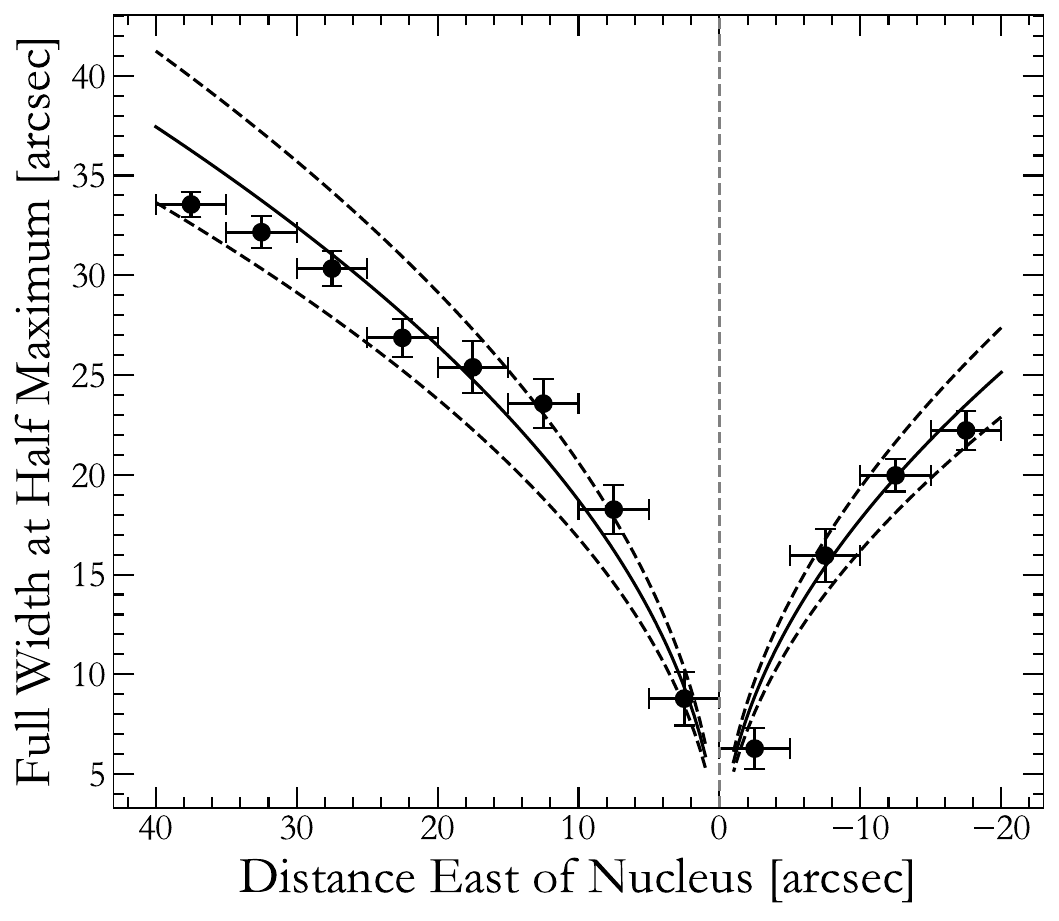}    \caption{FWHM of A3’s dust tail as a function of angular distance from the nucleus. Horizontal error bars denote the span of distances along the tail used in determining the FWHM, while vertical error bars reflect the measurement uncertainties in the FWHM profiles. The black solid lines indicate the best-fit relation, $\theta_{\perp} \propto l^{1/2}$, linking the tail’s FWHM to its distance from the nucleus. Derived dust ejection velocities are $v_{\perp}\sim(68\pm4)\,\beta^{1/2}$ m s$^{-1}$ to the east and $v_{\perp}\sim(61\pm5)\,\beta^{1/2}$ m s$^{-1}$ to the west.}
    \label{fwhm}
\end{figure}

An independent assessment of the size-velocity relation is obtained by examining the dust distribution in the direction perpendicular to the orbital plane. Notably, the observation on 13 April 2024 offers constraints on perpendicular distribution, as Earth was nearly coplanar with A3’s orbital plane at that time (\( \delta \sim -0.5^\circ \); see Table \ref{geometry}). Figure \ref{fwhm} shows the tail width $\theta_{\perp}$ as a function of the distance from the nucleus. The tail width is observed to remain below $38\arcsec$ for distances up to 40\arcsec east of the nucleus and below 25\arcsec for distances up to 20\arcsec west of the nucleus. Following \cite{jewitt2015episodic}, the ejection velocity perpendicular to the orbital plane, $v_{\perp}$, is determined from the measured tail width $\theta_{\perp}$ as follow
\begin{equation}
v_{\perp}=\left[\frac{\beta g_{\text{sun}}}{8 \ell }\right]^{1 / 2} \theta_{\perp},
        \label{PerVel}
\end{equation}
where $\ell$ denotes the projected distance from the nucleus, and $\theta_{\perp}$ and $\ell$ are expressed in meters. The data fitting shown in Figure \ref{fwhm} indicates that the relationship between $\theta_{\perp}$ and $\ell$ is given by $\theta_{\perp}=(6.2\pm0.3)\,\ell^{1/2}$ to the east of the nucleus and $\theta_{\perp}=(5.6\pm 0.4)\,\ell^{1/2}$ to the west of the nucleus. The dependence of $v_{\perp}$ on $\beta$ is determined employing Equation \ref{PerVel} along with the relation between $\theta_{\perp}$ and $\ell$. Specifically, to the east of the nucleus, the dust ejection velocity is $v_{\perp}\sim(68\pm4)\,\beta^{1/2}$ m s$^{-1}$, whereas to the west, it is $v_{\perp}\sim(61\pm5)\,\beta^{1/2}$ m s$^{-1}$. Given current uncertainties, the average dust ejection velocity is determined to be $v_{\perp}\sim(65\pm5)\,\beta^{1/2}$ m s$^{-1}$. This dependence is comparable with $v_{\mathrm{ej}}\sim (88\pm3)\,\beta^{1/2}$ m s$^{-1}$, as inferred from the surface brightness profiles of the coma (Equation \ref{vej}). The higher $v_{\mathrm{ej}}$ may result from an overestimated $l_{\text{coma}}$. If the dust brightness slope $q=-1$ is used as the criterion for the coma’s sunward extent, the $l_{\text{coma}}$ derived from the intersection of $q=-1$ and $q=-1.5$ slopes is likely overestimated, inflating $v_{\mathrm{ej}}$.

\subsection{Photometry}
\label{photometry}
\begin{figure}
\centering      \includegraphics[width=0.9\columnwidth]{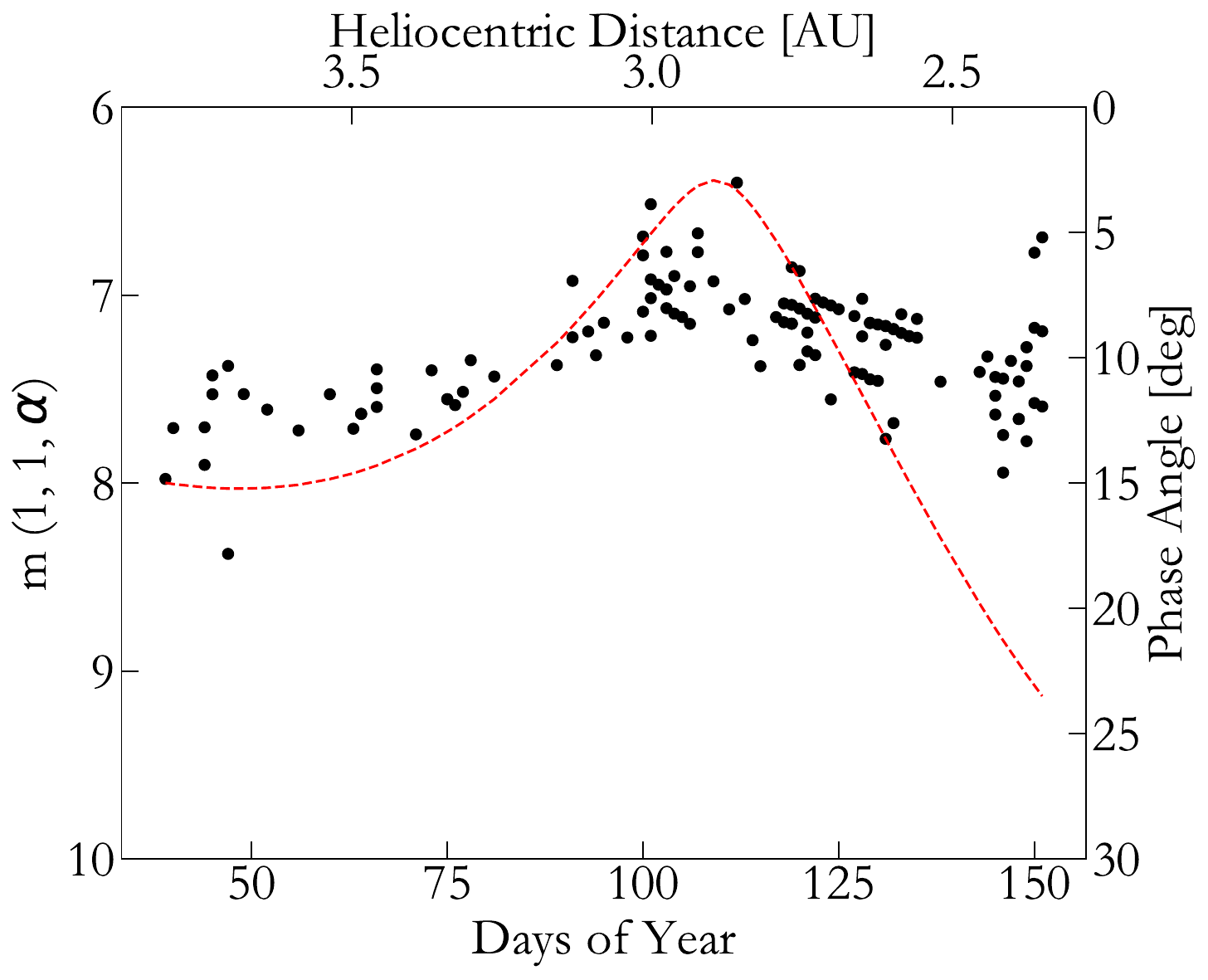}
    \caption{Reduced magnitude as a function of the observation date (Day 1 corresponding to 1 January 2024). The filled black circles represent the observational data collected by COBS, while the red dashed line illustrates the trend of the phase angle over time.}
    \label{PhaseAng_1}
\end{figure}

Photometry for each image is derived using circular apertures with radii between 10000 km and 160000 km. During the first observation, conducted at a geocentric distance of $\Delta = 3.6$ au, the angular radius of the smallest aperture is approximately 5\arcsec, which is about 1.5 times the FWHM of the point spread function. This choice of aperture size helps balance the need to capture the signal from A3's inner dust region while minimizing background noise, thereby optimizing the measurement of A3's dust properties. Additionally, to address contamination from field stars overlapping the apertures, their flux contributions were statistically estimated and subtracted using local background interpolation. The resulting apparent magnitudes $m_r(R, \Delta, \alpha)$ are reduced to absolute magnitudes $H$ by
\begin{equation}
H=m_r(R, \Delta, \alpha)-5 \log _{10}\left(R \Delta\right)-f(\alpha),
        \label{magnitude}
\end{equation}
where $R$ denotes the heliocentric distance, and $\Delta$ denotes the geocentric distance (refer to Table \ref{geometry}). It is assumed that the phase function $f(\alpha)$ varies linearly with the phase angle $\alpha$ such that $f(\alpha) = \gamma \alpha$. This linear function is determined from amateur observations, covering the period from 8 February 2024 to 30 May 2024, conducted by Comet Observation database (COBS\footnote{\href{https://www.cobs.si}{https://www.cobs.si}}). The magnitudes $m$ retrieved from amateur observations are corrected to the reduced magnitudes corresponding to geocentric and heliocentric distances of one au, and these reduced magnitudes $m\,(1, 1, \alpha)$ are presented in Figure \ref{PhaseAng_1}. From the figure, it is evident that the maximum value occurs on the $115_\text{th}$ day since 1 January 2024. Notably, the phase angle $\alpha$ of A3 on this day is also the smallest. This suggests the peak value can primarily be attributed to the brightness opposition effect; however, changes in cometary activity levels cannot be ruled out as a contributing factor. 

\begin{figure}
\centering      \includegraphics[width=0.9\columnwidth]{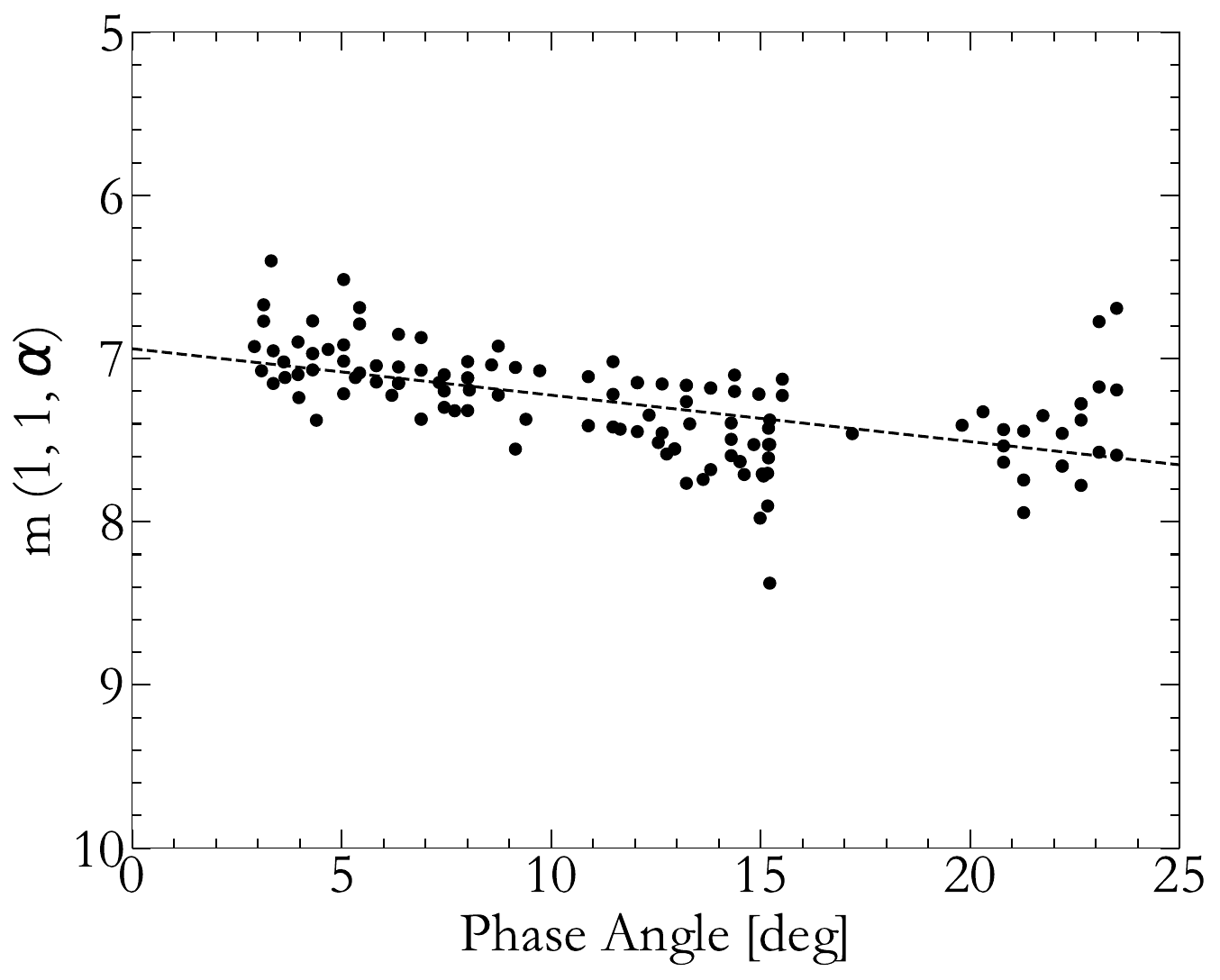}
    \caption{Reduced magnitude as a function of phase angle for A3. The black circles represent observational data from COBS. The dashed line denotes the best fit, exhibiting a gradient of $0.028 \pm 0.004$ mag deg$^{-1}$.}
    \label{PhaseAng}
\end{figure}

Figure \ref{PhaseAng} illustrates the relationship between $m\,(1, 1, \alpha)$ and $\alpha$, showing a general increase in brightness with decreasing phase angle. By fitting the data presented in Figure \ref{PhaseAng}, a linear coefficient $\gamma$ of ($0.028\pm0.004$) mag deg$^{-1}$ is obtained. This value falls within the typical range of $0.02$ to $0.04$ mag deg$^{-1}$ \citep{hui2019c,meech1988observations,bertini2019backscattering}. The phase function $f(\alpha)$ in Equation \ref{magnitude} is thus assumed to be $(0.028\pm0.004)\alpha$. The dust scattering cross-section, $C_e$ [km$^2$], is computed from the absolute magnitude $H$ by
\begin{equation}
C_e=\frac{2.24 \times 10^{16}\pi}{p_{r}}\times10^{0.4 (m_{\odot, r} - H)},
	\label{radius}
\end{equation}
where the solar $r$-band magnitude is adopted as $m_{\odot, r} = -26.9$ \citep{willmer2018absolute}. Following \citet{moreno2025dust}, we assume a geometric albedo of 0.04 for the cometary dust, which falls within the typical range of $0.03\sim0.06$\citep{hanner2003scattering}. 

The constraint on the radius of A3 is established through the nucleus photometry. Considering the seeing effects, a circular aperture with a constant radius of 10000 km, approximately 1.5 times the seeing FWHM, is employed for the nucleus photometry. To reduce the dust contamination in estimating the nucleus size, the background brightness is assessed using the median pixel intensity within an annular aperture extending from 10000 km to 15000 km. This configuration samples the background as close as possible to the photometric aperture, ensuring that the dust distribution in the background region closely matches that within the aperture, thereby subtracting the dust contribution by removing this background brightness. From the photometric data obtained on 23 February 2024, the measurement yields a minimum effective scattering cross-section of $C_e=(1.1\pm0.1)\times10^2$ km$^2$. The measured effective scattering cross-section corresponds to a radius of $5.9\pm0.2$ km for a homogeneous spherical body with a geometric albedo of 0.04, setting an upper limit on the radius of the nucleus. With an assumed nucleus density of 500 kg m$^{-3}$, the mass of the sphere is estimated to be $(4.3\pm0.8)\times10^{14}$ kg.

\begin{table*}
    \centering
    \renewcommand{\arraystretch}{1.4}
    \caption{Photometric results at different observation epochs}
    \label{PhotoTable}
    \setlength{\tabcolsep}{11pt}
    \begin{tabular}{lcccccc}
        \toprule
        Date & Property$^a$ & 10000 km & 20000 km & 40000 km & 80000 km & 160000 km \\
        \hline
23 Feb 2024 & $V$ & $14.03 \pm 0.02$ & $13.37 \pm 0.03$ & $12.91 \pm 0.05$ & $12.45 \pm 0.05$ & $12.36 \pm 0.06$ \\
23 Feb 2024 & $H$ & $8.26 \pm 0.06$ & $7.60 \pm 0.07$ & $7.14 \pm 0.08$ & $6.68 \pm 0.08$ & $6.59 \pm 0.09$ \\
23 Feb 2024 & $C_e$ & $6.1 \pm 0.2$ & $11.2 \pm 0.5$ & $17.1 \pm 0.9$ & $26.0 \pm 1.9$ & $28.3 \pm 2.3$ \\
23 Mar 2024 & $V$ & $13.05 \pm 0.04$ & $12.37 \pm 0.05$ & $11.82 \pm 0.05$ & $11.62 \pm 0.07$ & $11.38 \pm 0.07$ \\
23 Mar 2024 & $H$ & $8.23 \pm 0.06$ & $7.55 \pm 0.07$ & $7.00 \pm 0.07$ & $6.80 \pm 0.08$ & $6.56 \pm 0.08$ \\
23 Mar 2024 & $C_e$ & $6.2 \pm 0.4$ & $11.7 \pm 0.9$ & $19.4 \pm 1.4$ & $23.3 \pm 1.7$ & $29.0 \pm 1.8$ \\
13 Apr 2024 & $V$ & $12.12 \pm 0.07$ & $11.47 \pm 0.08$ & $10.89 \pm 0.11$ & $10.58 \pm 0.11$ & $10.48 \pm 0.09$ \\
13 Apr 2024 & $H$ & $8.10 \pm 0.08$ & $7.45 \pm 0.09$ & $6.87 \pm 0.13$ & $6.56 \pm 0.13$ & $6.46 \pm 0.11$ \\
13 Apr 2024 & $C_e$ & $7.0 \pm 0.7$ & $12.8 \pm 1.5$ & $21.8 \pm 3.2$ & $29.2 \pm 3.6$ & $31.9 \pm 3.2$ \\
7 May 2024 & $V$ & $11.86 \pm 0.06$ & $11.19 \pm 0.07$ & $10.75 \pm 0.09$ & $10.36 \pm 0.11$ & $10.27 \pm 0.12$ \\
7 May 2024 & $H$ & $8.15 \pm 0.08$ & $7.48 \pm 0.09$ & $7.04 \pm 0.10$ & $6.65 \pm 0.12$ & $6.56 \pm 0.13$ \\
7 May 2024 & $C_e$ & $6.7 \pm 0.5$ & $12.5 \pm 0.9$ & $18.7 \pm 1.7$ & $26.8 \pm 2.9$ & $29.2 \pm 3.9$ \\
30 May 2024 & $V$ & $11.92 \pm 0.07$ & $11.21 \pm 0.08$ & $10.71 \pm 0.09$ & $10.31 \pm 0.11$ & $10.15 \pm 0.13$ \\
30 May 2024 & $H$ & $8.13 \pm 0.12$ & $7.42 \pm 0.12$ & $6.92 \pm 0.13$ & $6.52 \pm 0.15$ & $6.36 \pm 0.16$ \\
30 May 2024 & $C_e$ & $6.8 \pm 0.7$ & $13.2 \pm 1.2$ & $20.8 \pm 2.1$ & $30.0 \pm 4.2$ & $34.9 \pm 5.3$ \\
        \hline
        \multicolumn{7}{l}{$^a$ Here, $V$ denotes apparent magnitude, $H$ denotes absolute magnitude, $C_e$ denotes effective scattering cross-section in 10$^3$ km$^2$.}\\ 
    \end{tabular}
\end{table*}

The dust properties of A3 across five observation epochs (from 23 February 2024 to 30 May 2024) are characterized by measurements obtained from projected apertures with radii ranging from 10000 km to 160000 km, as detailed in Table~\ref{PhotoTable}.
\begin{figure}
\centering	\includegraphics[width=0.9\columnwidth]{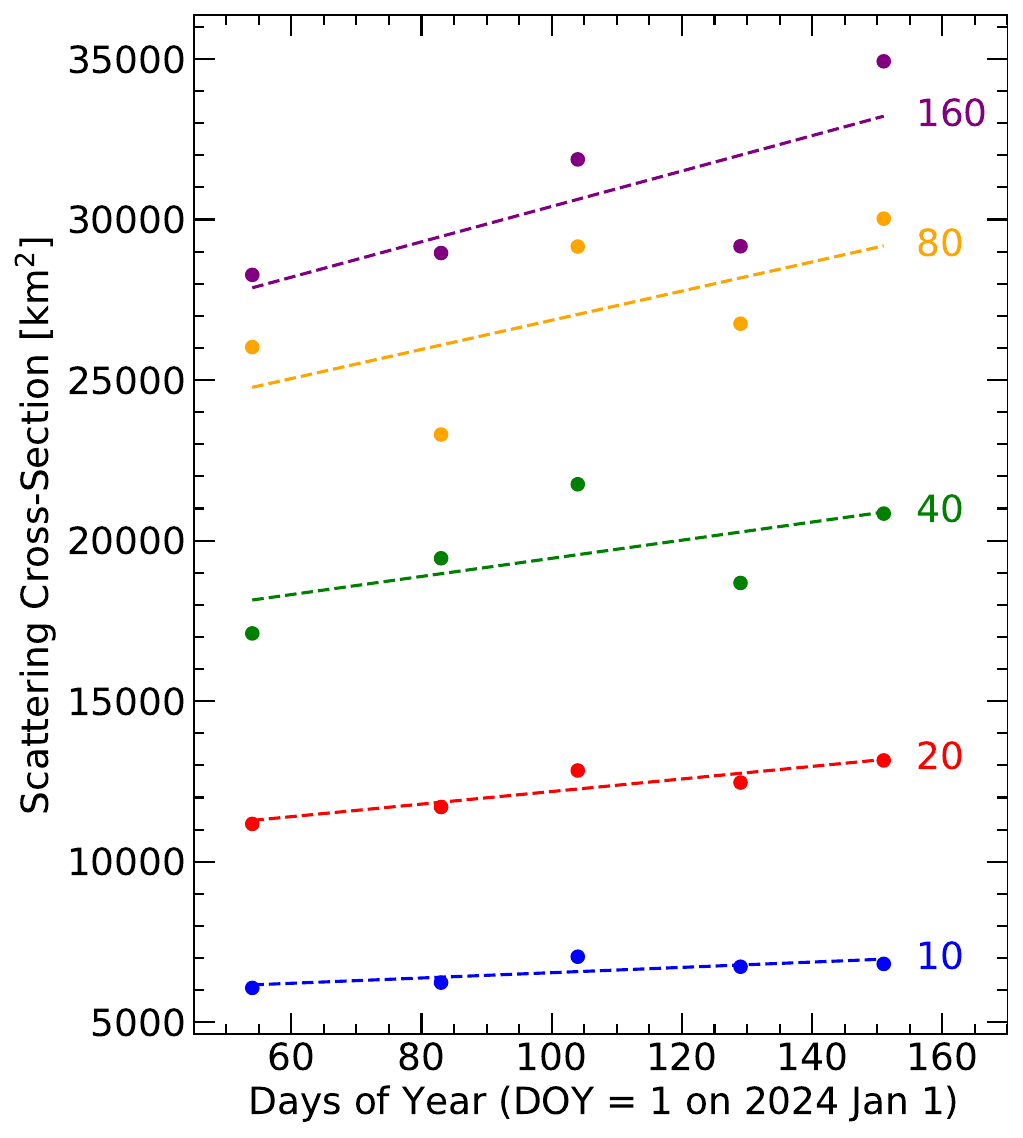}
    \caption{Effective scattering cross-section as a function of the observation date for different apertures. The radius of each aperture, expressed in units $10^3$ km, is indicated next to the corresponding line. The dashed lines represent linear fits to the data points, with coefficients of determination ($R^2$) of 0.59, 0.83, 0.35, 0.42, and 0.56 for the fits corresponding to apertures from smallest to largest, respectively.}
    \label{area}
\end{figure}

\section{Discussion}
\label{discussion}
\subsection{Dust properties}
\label{Dust properties}
The variation of the effective scattering cross-section within different apertures (shown in Table \ref{PhotoTable}) as a function of time is illustrated in Figure \ref{area}. Examination of Figure \ref{area} reveals that the effective scattering cross-section exhibits a generally steady and gradual increase over the observational period, indicating that A3’s dust activity level has evolved in a relatively stable and moderate manner. This relies on a linear phase function fitted to the COBS dataset (coefficient $0.028\pm0.004$ mag deg$^{-1}$) applied to correct the ZTF measurements, which span the same period but are sparser. If the COBS data include activity surges not sampled by ZTF, the phase correction might overestimate geometric effects, potentially flattening the trend in Figure \ref{area}. To evaluate this, we refitted the phase function using COBS observations aligned with ZTF dates, obtaining $0.022\pm0.006$ mag deg$^{-1}$. The close agreement with the original value indicates that the trend shown in Figure \ref{area} reliably reflects the dust activity level. The peak observed in the effective scattering cross-section on April 13 is primarily attributed to the smaller phase angle at that time. 

A linear regression is performed on the data from all apertures, as depicted in Figure \ref{area}, using the linear equation $C=C_0+kt$, where $C$ represents the effective scattering cross-section at a given time, $C_0$ denotes the effective scattering cross-section at DOY = 0 (corresponding to 31 December 2023), $k$ represents the rate of change of the effective scattering cross‑section over time, and $t$ is the observation time. Using the parameters $C_0$ and $k$ derived from the linear fit, the initial time $t_0$ at which the effective scattering cross-section would be zero is estimated, which approximates the onset time of dust activity.
\begin{table*}[ht]
    \centering
    \renewcommand{\arraystretch}{1.4}
    \caption{Fixed-aperture Photometry vs.~Date}
    \setlength{\tabcolsep}{16pt}
    \begin{tabular}{lcccccc}
    \toprule
    Radius$^{a}$ & $C_0$ $^{b}$ & $k$ $^{c}$ &$s$ $^d$ & $t_0$ $^{e}$ & $T_0$ $^{f}$ & $R_0$ $^{g}$  \\
\hline
    $10$   & $5.7\pm0.4$   & $8.3 \pm 4.0$   & 0.3& $-692 \pm 340$   &  7 Feb 2022   & 10.4 \\
    $20$  & $10.2\pm0.6$   & $19.6 \pm 5.1$  &0.4 & $-523 \pm 139$   & 26 Jul 2022   & 9.1\\
    $40$ & $16.6\pm2.4$  & $28.2 \pm 22.3$   &1.7& $-589 \pm 474$   & 22 May 2022   & 9.6 \\
    $80$ & $22.3\pm3.4$  & $45.4 \pm 30.9$  &2.3& $-491 \pm 342$   & 6 Sep 2022   & 8.7 \\
    $160$ & $24.9\pm3.0$ & $55.1 \pm 27.3$ &2.1& $-452 \pm 231$   & 5 Oct 2022   & 8.5\\
\hline
Weighted Mean& & & & $-524\pm104$ & 25 Jul 2022 & 9.1\\
\hline
\multicolumn{7}{p{0.9\textwidth}}{$^{a}$Projected radius of the photometry aperture, in units of $10^3$ km. $^{b}$Effective scattering cross-section at DOY = 0, in units of $10^3$ km$^2$. $^{c}$Rate of change of the scattering cross-section, in square kilometers per day. $^{d}$Standard deviation of residuals, in units of $10^3$ km$^2$. $^{e}$Initial time expressed as days prior to 31 December 2023. $^{f}$Initiation time without the uncertainty expressed in standard date format. $^{g}$Heliocentric distance at time $T_0$, in astronomical units.}
\end{tabular}
\label{tab:photometry_vs_time}
\end{table*}

The optimal values for $C_0$, $k$ and $t_0$ corresponding to each aperture, along with the calculated heliocentric distances $R_0$ at the onset of activity, are summarized in Table \ref{tab:photometry_vs_time}. The mean onset time of dust activity is approximately $-524\pm104$ days, corresponding to 25 July 2022 (within the uncertainties). At this time, the heliocentric distance of A3 is approximately 9.1 au. The lack of data preceding the observational period precludes empirical validation of the consistency between the scattering cross-section's rate of change and the derived linear trend. Therefore, the extrapolated onset time of activity should be considered an estimate and interpreted with caution. It is noteworthy that the inferred onset time of dust activity is approximately five months prior to the earliest observational records, suggesting the potential plausibility of the inference. Furthermore, the estimated activation heliocentric distance exceeds the critical sublimation distance for water ice, which is approximately $R\sim4-5$ au \citep{jewitt2019initial}. This observation implies that water ice sublimation alone may not fully explain the initiation of dust activity at such a distant location, whereas sublimation of CO or CO$_2$ typically occurs beyond 10 au due to its lower sublimation temperature, so it seems too distant to effectively explain the activity of A3. In this context, the phase transition from amorphous to crystalline ice is likely to play a significant role, as it typically occurs within 5–17 au range \citep{korsun2003dust}. The transition not only releases energy but also liberates CO or CO$_2$ trapped within the amorphous ice, thereby triggering dust loss from the nucleus \citep{korsun2003dust}. Since A3's initial heliocentric distance falls well within this range, this indicates that the crystallization of amorphous ice may be the primary driver of its dust activity.

An estimate of upper limit of the minimum size of dust particles can be derived from the length of the tail, which is about $l\sim30\arcsec$ ($\sim70000$ km) on 23 February 2024. Under the influence of solar radiation pressure, a decreasing heliocentric distance enables the production of smaller dust particles that can travel farther. Assuming the tail's edge comprises the smallest particles ejected at various times, the size of dust particles emitted at the onset of activity and reaching the tail's edge provides an upper limit for the minimum dust particle size. The size of dust particles capable of traveling a distance $l$ within a time $t\sim5.2\times10^7$ s ($\sim 600$ days) is determined to be
\begin{equation}
\beta=\frac{2l}{g_{\text{sun}}t^2}.
	\label{min_size}
\end{equation}
The minimum radius of dust particles is determined by the maximum dimensionless factor $\beta_\text{max}$, which is estimated to be approximately $7\times10^{-4}$. The factor $\beta$ is defined as $\beta = C_{\mathrm{pr}} Q_{\mathrm{pr}}/(2 \rho a)$. In this expression, $a$ denotes the particle's radius and $\rho$ is its bulk density, assumed here to be $\rho = 500\, \text{kg/m}^3$. Following \cite{liu2024unraveling}, the solar radiation pressure constant $C_{\mathrm{pr}}$ is set to $1.19\times10^{-3}$ m s$^{-2}$ and $Q_{\mathrm{pr}}$ is assigned a value of 1. Therefore, the relationship between the dust particle radius $a$ and $\beta$ can be simplified as $a\sim1/\beta$, and the upper limit of the minimum radius of dust particles is estimated to be about 1.4 mm. This upper limit provides a constraint for the minimum radius of the particles in subsequent dust dynamical simulations. 

\subsection{Dust dynamics simulation}
\label{Tail Dynamical Modelling}
The brightness distribution surrounding the nucleus in observational images of A3 is influenced by the physical properties and dynamical behavior of the ejected dust particles. By integrating numerical simulations with observational data, it becomes possible to place constraints on the information related to the dust ejection activities, such as the particle's emission time, initial position, initial velocity, and grain size. 

A dust dynamics simulation procedure, based on the framework developed by \cite{liu2016dynamics} and further refined by \cite{liu2024unraveling} and \cite{liu2025physical}, is adapted for this study and used to generate the simulated image of A3. In the model, the motion of dust particles emitted from the nucleus is governed by the influences of solar gravity and solar radiation pressure. The equations of motion of dust particles, formulated within the heliocentric inertial reference frame, are numerically integrated using the Gragg-Bulirsch-Stoer (GBS) method \citep{stoer1980introduction} with adaptive stepsize, offering a good compromise between speed and accuracy. Through this integration process, the spatial position of the particle at any given observational epoch is determined. Subsequently, this spatial position is projected onto the observer’s sky plane via coordinate transformation. In this plane coordinate system, the spatial scale of grid cells is aligned with the resolution of the observational imaging system. The contribution of each dust particle to the brightness of the corresponding grid cell is then quantitatively evaluated, thereby enabling the simulation of cometary observational images. 

During the simulation of observation images, model parameters that exert a marked influence on the results are primarily taken into account. These parameters include the activity duration ($t_1$ and $t_0$), the ejection velocity ($v_\text{ej}$), the dust particle size range ($\beta_\text{max}$ and $\beta_\text{min}$), and the size distribution index ($s$). In order to reduce computational load, the input parameters are constrained prior to running the simulation. The activity's onset time $t_1$ has previously been determined through a fitting of the scattering cross-section's data extrapolation, yielding an approximate date of 10 July 2022. Moreover, the rise in the effective scattering cross-section shown in Figure \ref{area} indicates that dust release is continuing, so the terminal time, $t_0$, has been set to the observation date. The ejection velocity is assumed to follow the relation $v_\text{ej} = v_0 \beta^{\gamma}$, with $v_0$ having been determined as 65 m/s, as described in Section \ref{morphology}. The maximum dust size ($\beta_\text{min}$) is derived from the simple physical condition by equating the gas drag force to the gravitational force acting on the dust grain \citep{jewitt2019distant}. In this condition, the following equation is obtained: 
\begin{equation}
\beta_\text{min} = \frac{2 \pi G \rho^2 r_\text{n}}{1125C_\text{D} V_\text{g} f_\text{s}},
	\label{a_max}
\end{equation}
where $G$ is the gravitational constant, $\rho$ represents the bulk density of both the nucleus and the dust, assumed to be 500 kg m$^{-3}$, $r_\text{n}=5.9\pm0.2$ km is the radius of the nucleus (used in kilometers), $C_\text{D}=1$ is the dimensionless drag coefficient, $V_\text{g}=500$ m s$^{-1}$ is the assumed gas speed at the surface of the nucleus, $f_\text{s}$ is the mass flux of the sublimated ice, which is determined later in this paper to be approximately $\sim10^{-5}$ kg m$^{-2}$ s$^{-1}$. Substitution of the defined parameters yields $\beta_\text{min}=0.0001$, with a corresponding maximum dust grain size of approximately 10 mm. The remaining parameters, the particle size distribution index and minimum radius of the particles, are treated as free variables within this model.

\begin{figure}
\centering
        \includegraphics[width=1\columnwidth]{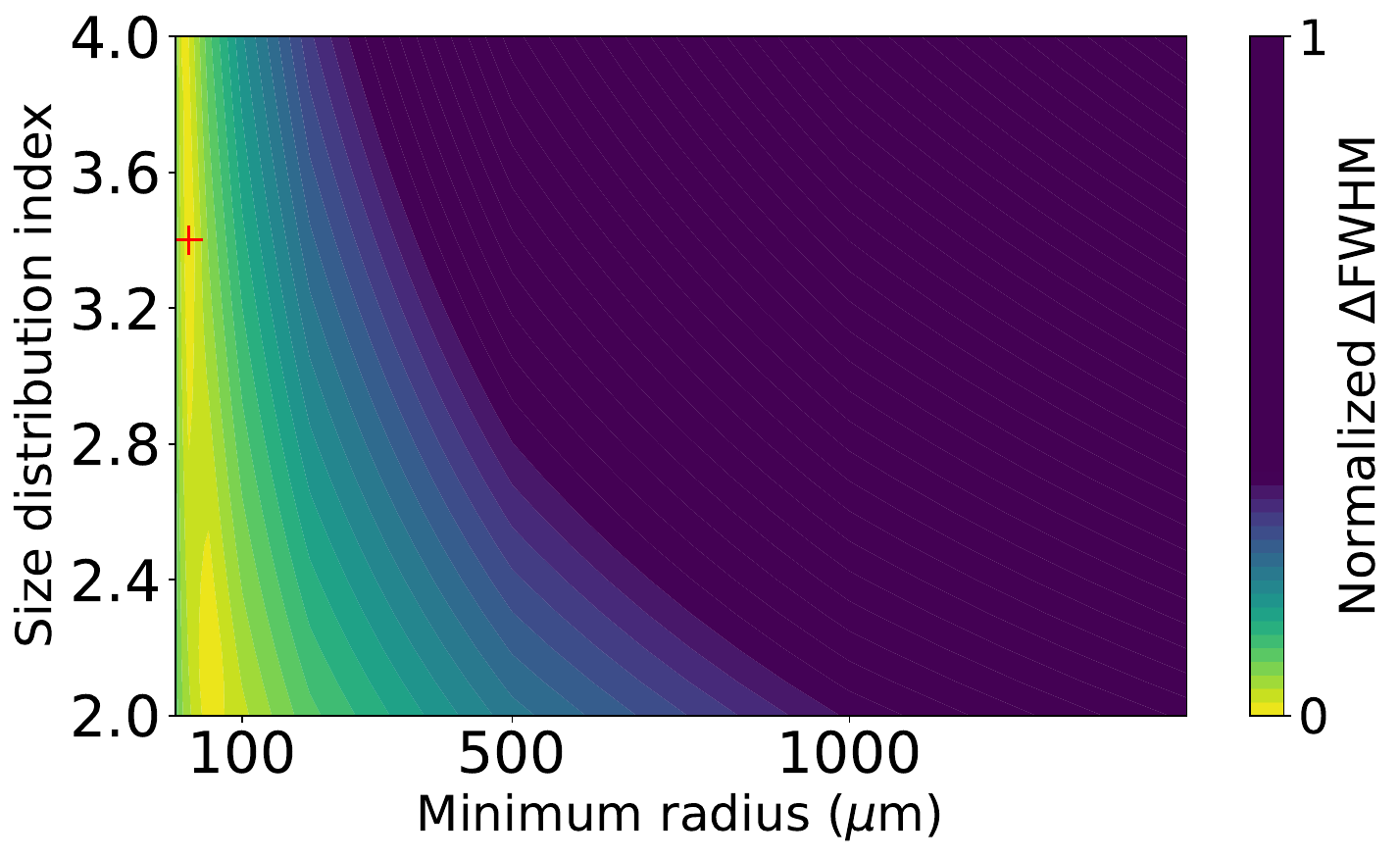}
    \caption{Contour plot of the normalized $\Delta \text{FWHM}$ as a function of minimum particle radius and size distribution index. The red cross marks the best-fit parameters ($a_{\text{min}} = \SI{20}{\micro\meter}$, $s = 3.4$).}
    \label{modelsu}
\end{figure}

\begin{figure}
\centering
	\includegraphics[width=0.9\columnwidth]{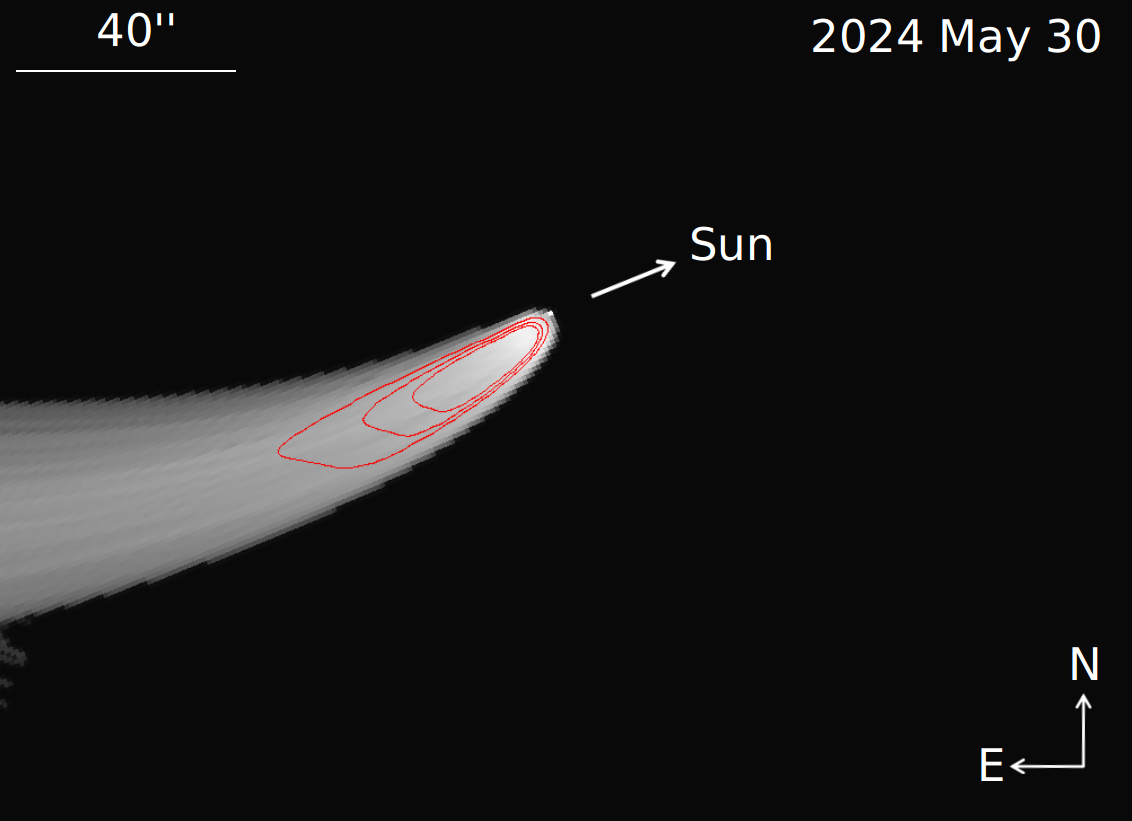}
    \caption{The modeled morphology of A3. True north (N) and true east (E) directions are denoted by white arrows, with the direction toward the Sun also indicated. A scale bar is provided for reference. Contours represent isophotes at brightness levels of 1/2, 1/4, and 1/8 of the peak brightness, consistent with the step size used in the observed image.}
    \label{dustmodel}
\end{figure}

Simulations are conducted by varying the particle size distribution index, $s$, from 2.0 to 4.0 in increments of 0.1, and by selecting 10 different minimum radii ranging from $\SI{1}{\micro\meter}$ to 1.5 mm. For each simulation, the FWHM of the radial profile is computed and compared with the observed FWHM from the 30 May 2024 observation to determine the best-fit parameters. The absolute difference between the simulated and observed FWHM, $\Delta \text{FWHM} = |\text{FWHM}_{\text{sim}}- \text{FWHM}_{\text{obs}}|$, is calculated and normalized using Min-Max  normalization. For visualization, a contour plot of the normalized $\Delta\text{FWHM}$ is shown in Figure~\ref{modelsu}, with the best-fit parameters marked by a red cross. We prefer not to overinterpret the result. The obtained best-fit parameters are $a_{\text{min}} = \SI{20}{\micro\meter}$ and $s = \num{3.4}$. Deviations from these values, result in less satisfactory matches. The result of the size distribution power index of 3.4 before perihelion, with a minimum particle size of \SI{20}{\micro\meter}, aligns with the high polarization degrees reported by \cite{lim2025optical}, indicating a notable contribution of small, highly scattering dust particles. The simulated image, created with this parameter set and convolved with a Gaussian function having a FWHM adapted to the seeing conditions, is shown in Figure \ref{dustmodel}. The position angle of the linear feature in this image, measured as $107.5\pm0.3\degr$, almost agrees with the $109.9\pm0.2\degr$ determined for the corresponding feature in the observed image from 30 May 2024.

The the average size achieved by combining the size distribution index with the maximum and minimum sizes previously obtained is computed as follows:
\begin{equation}
\bar{a}=\left( \frac{3 - s}{4 - s} \right)
\left( \frac{a_{\text{max}}^{4 - s} - a_{\text{min}}^{4 - s}}{a_{\text{max}}^{3 - s} - a_{\text{min}}^{3 - s}} \right).
\label{a_aver}
\end{equation}
Substituting the previously obtained values of $a_{\text{max}}=10$ mm and $a_{\text{min}}=\SI{20}{\micro\meter}$ into the equation, the average radius of the particles is estimated to be about 0.4 mm. Table \ref{tab:photometry_vs_time} shows a rate of change of the effective scattering cross‑section over time of $\frac{ \mathrm{d} C_\text{e}}{ \mathrm{d}t} = 55\pm27$ km$^2$ day$^{-1}$ within a circular aperture spanning a radius of 160000 km. The corresponding dust production rate is determined by 
\begin{equation}
\frac{\mathrm{d}M_\mathrm{d}}{\mathrm{d}t}=\frac{4\bar{a}\rho}{3}\frac{ \mathrm{d} C_\text{e}}{ \mathrm{d}t},
\label{mass_production}
\end{equation}
where $\rho$ represents the bulk mass density of dust, assumed to be the same as the cometary nucleus bulk density (500 kg/m$^3$). By substituting the assumed and obtained values into Equation \ref{mass_production}, a dust production rate of approximately $(1.7 \pm 0.8) \times 10^2$ kg s$^{-1}$ is determined. Assuming the dust production rate has remained consistent since the onset of dust activity, it can be inferred that by the time of the last observation, a total dust mass of approximately $10^{10}$ kg would have been lost from the nucleus. This corresponds to approximately $10^{-4}$ of the nucleus’s total mass, assuming a previously estimated radius of $6$ km.

Our analysis of A3’s dust loss, based on pre-perihelion imaging from ZTF spanning 3.6 au to 2.4 au, reveals an average dust mass loss rate of 170 kg/s before perihelion, with the particle size distribution index of s = 3.4 and a minimum particle size of about \SI{20}{\micro\meter}. Examination of the dust distribution perpendicular to the orbital plane reveals that the relationship of the ejection velocity and the dust size is $v_{\perp}\sim(65\pm5)\,\beta^{1/2}$ m s$^{-1}$. The initial time of significant dust loss, derived from extrapolation of the scattering cross-section trend, is estimated to be 25 July 2022. Recently, \cite{moreno2025dust} conducted a detailed modeling of the dust environment of C/2023 A3 across its pre- and post-perihelion phases. For the same orbital segment, our derived velocities and production rates align well with their findings. Besides, \cite{moreno2025dust} report a dust size distribution index ranging from 3.5 to 3.9, slightly higher than our value of 3.4. They also determined a dust size range decreasing from \SI{10}{\micro\meter} pre-perihelion to \SI{1}{\micro\meter} near perihelion, which is smaller compared to our results.

%0226
\subsection{Stability of the nucleus}
Long-period comets experience the combined effects of solar heating and tidal forces as they approach the Sun, which contribute to the structural instability of the nucleus and potentially trigger its disintegration. \cite{sekanina20191i} reported that comets originating from the Oort Cloud with perihelion distances of less than 0.6 au generally face a higher risk of disintegration as they near the Sun. A statistical analysis of a subset of long-period comets suggests that the median perihelion distance for those that fragment is approximately 0.49 au \citep{jewitt2022destruction}. Notably, A3's perihelion distance (0.39 au) is lower than this median, suggesting that it is at a statistically significant risk of disintegration. Previous studies have indicated that comet A3 may undergo fragmentation as it approaches perihelion \citep{sekanina2024inevitable}. This section will examine the potential for A3's disintegration by considering several primary mechanisms that could drive such an event, including tidal effect, sublimation-induced erosion, and rotational instability.

To assess the potential for tidal forces to contribute to the fragmentation of the nucleus, the tidal effects exerted by the Sun are examined, and an estimate of the tidal effect on the surface on the nucleus is performed using the dimensionless parameter $\epsilon$:
\begin{equation}
\epsilon=\frac{a_\text{t}}{a_\text{g}}.
\label{tidal effect}
\end{equation}
The tidal acceleration $a_{\text{t}}$ is expressed as $a_\text{t}=\frac{2GM_\odot r_\text{n}}{R^3}$, where $M_\odot=1.989\times10^{30}$ kg is the mass of the Sun. The
self-gravitational acceleration at the surface of the nucleus $a_\text{g}$ is expressed as $a_\text{g}=\frac{4}{3}\pi G\rho r_\text{n}$. When the dimensionless parameter $\epsilon>1$, the nucleus is highly susceptible to tidal disruption. It is found that at perihelion, the dimensionless parameter $\epsilon$ is about $10^{-5}$. Therefore, it can be inferred that the tidal forces exerted by the Sun are insufficient to overcome the gravitational binding of A3's nucleus, effectively ruling out the possibility of tidal disintegration.

Sublimation-induced erosion may lead to the disintegration of a nucleus. To evaluate this, the mass flux of the sublimated ice $F_s(T)$ is first derived using the radiation balance thermal equation, given by:
\begin{equation}
\frac{F_{\odot}}{R^{2}}(1-A)=\chi\left[\epsilon \sigma T\left(R\right)^{4}+F_{s} L_{s}(T)\right],
        \label{waterRate}
\end{equation}
where $F_{\odot}=1361$ W m$^{-2}$ corresponds to solar constant, $A = 0.1$ represents the assumed albedo, $\sigma = 5.67\times10^{-8}$ W m$^{-2}$ K$^{-4}$ denotes the Stefan–Boltzmann constant, $L_s (T)$ indicates the latent heat of water ice at the temperature $T$, and $\epsilon = 0.9$ is the assumed effective emissivity. The dimensionless parameter $\chi$ is defined as the ratio between the effective solar heat-absorbing area and the thermal radiating area of a cometary nucleus. This parameter ranges from $\chi = 1$, corresponding to a non-rotating nucleus with a subsolar ice patch, to $\chi = 4$, representing an idealized spherical nucleus with uniform temperature (isothermal). A value of $\chi = 2$ is applied as an intermediate case, representing a nucleus where the absorption and emission are both confined to one hemisphere. Following \cite{jewitt2023disintegration}, Equation \ref{waterRate} is then used to estimate the rate of loss of nucleus material by $\text{d}r_\text{n}/\text{d}t=-F_\text{s}/\rho$. With the $F_s\sim1.3\times10^{-4}$ kg m$^{-2}$ s$^{-1}$ at 1 au from Equation \ref{waterRate}, the erosion rate of the nucleus due to sublimation is derived as about $-8.5\,\mathrm{m\,year}^{-1}$. At this erosion rate, it would take approximately 700 years for the entire nucleus to erode. In comparison, the time spent within the snow line (where $R<3$ au for the minimum-mass solar nebula model from \citet{okuzumi2012rapid}) is on the order of 1 year. Therefore, from this perspective, sublimation is unlikely to lead to the disintegration of A3 through stable erosion alone.

As a common cause of nucleus disintegration, rotational instability is also an important mechanism to investigate. The main factor that could contribute to such instability is the torque induced by the sublimation. Sublimation leads to asymmetric outgassing, which in turn applies torques to the nucleus, potentially altering its rotational state. To quantify the extent of this effect, it is important to evaluate the time scale $\tau_\text{s}$ required for sublimation-induced torques to cause a significant change in the nucleus’s rotation. This time scale, as derived by \cite{jewitt2022destruction}, is given by:
\begin{equation}
\tau_\text{s} = \left( \frac{16\pi^2}{15} \right) \left( \frac{\rho r_\text{n}^4}{k_\text{T} V_\text{g} P} \right) \left( \frac{1}{\dot{M}} \right),
\label{rotationalInsta}
\end{equation}
where $k_\text{T}$ represents the dimensionless moment arm, $P$ is the rotational period, and $\dot{M}$ is the mass loss rate of the nucleus. Following \cite{jewitt2021systematics}, $k_\text{T}$ and $P$ are adopted as 0.007 and 15 hours, respectively. The parameter $\dot{M}$ is calculated based on the assumption that A3 contains abundant water ice, and that rapid sublimation occurs only for heliocentric distances $R < 3$ au. The production rate of water was determined as $Q_{H_{2}O}=(1.50\pm0.37)\times10^{28}$ s$^{-1}$ ($\sim(4.5\pm1.1)\times10^2$ kg s$^{-1}$) from pre-perihelion observation on 31 May 2024 at $R= 2.36$ au \citep{ahuja2024molecular}. With $F_\text{s}\sim2.4\times10^{-5}$ kg m$^{-2}$ s$^{-1}$ at $R= 2.36$ au from Equation \ref{waterRate}, the area responsible for maintaining the sublimation activity is estimated to be $A_\text{s}\sim 2\times10^7$ m$^2$. The calculated active area is equivalent to the surface of a sphere with an estimated radius of 1.26 km. Assuming that the sublimation activity is confined within this area, and considering the strong correlation between the nucleus' activity and its heliocentric distance, it follows from Equation \ref{waterRate} that the mass loss rate of the nucleus peaks at perihelion, reaching up to $\dot{M}_\text{max}\sim20000$ kg s$^{-1}$. Given $\rho$ = 500 kg m$^{-3}$, $r_\text{n} = 6$ km, and $V_\text{g}=500$ m s$^{-1}$, the lower limit of $\tau_s$ is derived to be approximately $1.8\times10^9$ s ($\sim50$ years). 

The derived lower limit of $\tau_s$ is much greater than the duration $\Delta t$ ($\sim1$ year) that A3 spends within 3 au of the Sun, indicating that the sublimation-induced torque is too weak to cause a significant change in the nucleus' rotational state during a single passage. Therefore, the rotational instability driven by sublimation would not be sufficient to lead to fragmentation of the nucleus. Having ruled out three primary mechanisms that could lead to the fragmentation of the nucleus, it is concluded that the nucleus is unlikely to fragment near perihelion, consistent with subsequent observations which show that the nucleus has survived.

\section{Conclusions}
\label{CONCLUSIONS}
The dust loss of the Great Comet C/2023 A3 (Tsuchinshan–ATLAS) during the period from 23 February 2024 to 30 May 2024 is analysed in this study. The results are listed as follows:

1. The dust loss from A3 occurs continuously and remains in a steady state, as indicated by the logarithmic surface brightness slopes of the inner coma (within the range of 7\arcsec) being close to $q$ = -1. The velocity component perpendicular to the orbital plane, $v_{\perp}$, exhibits a dependence on dust size $\beta$, which is described by the relation $v_{\perp}\sim(65\pm5)\,\beta^{1/2}$ m s$^{-1}$.

2. The photometry reveals that the upper limit of the radius of the nucleus is estimated to be $5.9\pm0.2$ km with an assumed albedo of 0.04. Extrapolation of the trend in the dust scattering cross-section indicates that the dust activity started on 25 July 2022 (within the uncertainties), at which time A3's heliocentric distance is 9.1 au. The mechanism associated with this initial distance is the phase transition from amorphous to crystalline ice.

3. The dynamics simulation of the dust tail indicates that the size range of dust particles in the observational images is approximately \SI{20}{\micro\meter} to 10 mm, with a size distribution index of $s = 3.4$. The dust loss rate is determined to be about $(1.7 \pm 0.8) \times 10^2$ kg s$^{-1}$.

4. Analysis of the stability of the nucleus indicates that tidal effects, sublimation, and rotational instability are unlikely to lead to the disintegration of the nucleus, consistent with the observational results showing that the nucleus has survived.

\begin{acknowledgements}
We thank the editor and the anonymous reviewer for their insightful comments on our work. The work is supported by the Science and Technology Development Fund (FDCT) of Macau (grant No. 0016/2022/A1) and the Faculty Research Grants of the Macau University of Science and Technology (grant No. FRG-25-035-SSI) to M.T.H, and the National Natural Science Foundation of China (No.~12472048) to X. Liu.
\end{acknowledgements}
%%%%%%%%%%%%%%%%%%%%%%%%%%%%%%%%%%%%%%%%%%%%%%%%%%.

\bibliography{aanda}{}

\begin{thebibliography}{36}
\expandafter\ifx\csname natexlab\endcsname\relax\def\natexlab#1{#1}\fi

\bibitem[{Ahuja {et~al.}(2024)Ahuja, Aravind, Sahu, Jehin, Donckt, Hmiddouch, Ganesh, \& Sivarani}]{ahuja2024molecular}
Ahuja, G., Aravind, K., Sahu, D., {et~al.} 2024, The Astronomer's Telegram, 16637, 1

\bibitem[{Bailey(1996)}]{bailey1996provenance}
Bailey, M. 1996, Earth, Moon, and Planets, 72, 57

\bibitem[{Bertini {et~al.}(2019)Bertini, La~Forgia, Fulle, Tubiana, G{\"u}ttler, Moreno, Agarwal, Mu{\~n}oz, Mottola, Ivanovsky, {et~al.}}]{bertini2019backscattering}
Bertini, I., La~Forgia, F., Fulle, M., {et~al.} 2019, Monthly Notices of the Royal Astronomical Society, 482, 2924

\bibitem[{Betzler \& Borges(2012)}]{betzler2012nonextensive}
Betzler, A.~S. \& Borges, E.~P. 2012, Astronomy \& Astrophysics, 539, A158

\bibitem[{Dones {et~al.}(2004)Dones, Weissman, Levison, \& Duncan}]{dones2004oort}
Dones, L., Weissman, P.~R., Levison, H.~F., \& Duncan, M.~J. 2004, in Star Formation in the Interstellar Medium: In Honor of David Hollenbach, Vol. 323, 371

\bibitem[{Graham {et~al.}(2019)Graham, Kulkarni, Bellm, Adams, Barbarino, Blagorodnova, Bodewits, Bolin, Brady, Cenko, {et~al.}}]{graham2019zwicky}
Graham, M.~J., Kulkarni, S., Bellm, E.~C., {et~al.} 2019, Publications of the Astronomical Society of the Pacific, 131, 078001

\bibitem[{Grant \& Jones(2024)}]{grant2024prospects}
Grant, S.~R. \& Jones, G.~H. 2024, Research Notes of the AAS, 8, 252

\bibitem[{Hands \& Dehnen(2020)}]{hands2020capture}
Hands, T. \& Dehnen, W. 2020, Monthly Notices of the Royal Astronomical Society: Letters, 493, L59

\bibitem[{Hanner(2003)}]{hanner2003scattering}
Hanner, M.~S. 2003, Journal of Quantitative Spectroscopy and Radiative Transfer, 79, 695

\bibitem[{Hui {et~al.}(2019)Hui, Farnocchia, \& Micheli}]{hui2019c}
Hui, M.-T., Farnocchia, D., \& Micheli, M. 2019, The Astronomical Journal, 157, 162

\bibitem[{Jehin {et~al.}(2024)Jehin, Donckt, Hmiddouch, \& Manfroid}]{jehin2024trappist}
Jehin, E., Donckt, M.~V., Hmiddouch, S., \& Manfroid, J. 2024, The Astronomer's Telegram, 16705, 1

\bibitem[{Jewitt(2021)}]{jewitt2021systematics}
Jewitt, D. 2021, The Astronomical Journal, 161, 261

\bibitem[{Jewitt(2022)}]{jewitt2022destruction}
Jewitt, D. 2022, The Astronomical Journal, 164, 158

\bibitem[{Jewitt {et~al.}(2019)Jewitt, Agarwal, Hui, Li, Mutchler, \& Weaver}]{jewitt2019distant}
Jewitt, D., Agarwal, J., Hui, M.-T., {et~al.} 2019, The Astronomical Journal, 157, 65

\bibitem[{Jewitt {et~al.}(2015)Jewitt, Agarwal, Weaver, Mutchler, \& Larson}]{jewitt2015episodic}
Jewitt, D., Agarwal, J., Weaver, H., Mutchler, M., \& Larson, S. 2015, The Astrophysical Journal, 798, 109

\bibitem[{Jewitt {et~al.}(2023)Jewitt, Kim, Mattiazzo, Mutchler, Li, \& Agarwal}]{jewitt2023disintegration}
Jewitt, D., Kim, Y., Mattiazzo, M., {et~al.} 2023, The Astronomical Journal, 165, 122

\bibitem[{Jewitt \& Luu(2019)}]{jewitt2019initial}
Jewitt, D. \& Luu, J. 2019, The Astrophysical Journal Letters, 886, L29

\bibitem[{Jewitt \& Meech(1987)}]{jewitt1987surface}
Jewitt, D. \& Meech, K.~J. 1987, Astrophysical Journal, Part 1 (ISSN 0004-637X), vol. 317, June 15, 1987, p. 992-1001. NASA-supported research., 317, 992

\bibitem[{Joye \& Mandel(2003)}]{joye2003new}
Joye, W.~A. \& Mandel, E. 2003, in Astronomical data analysis software and systems XII, Vol. 295, 489

\bibitem[{Kim {et~al.}(2020)Kim, Jewitt, Mutchler, Agarwal, Hui, \& Weaver}]{kim2020coma}
Kim, Y., Jewitt, D., Mutchler, M., {et~al.} 2020, The Astrophysical Journal Letters, 895, L34

\bibitem[{Korsun \& Ch{\"o}rny(2003)}]{korsun2003dust}
Korsun, P. \& Ch{\"o}rny, G. 2003, Astronomy \& Astrophysics, 410, 1029

\bibitem[{Lim {et~al.}(2025)Lim, Ishiguro, Takahashi, Akitakya, Geem, Bach, Jin, Jo, Choi, Seo, {et~al.}}]{lim2025optical}
Lim, B., Ishiguro, M., Takahashi, J., {et~al.} 2025, The Astrophysical Journal Letters, 983, L19

\bibitem[{Liu {et~al.}(2025)Liu, Li, Mu, \& Liu}]{liu2025physical}
Liu, B., Li, C., Mu, Z., \& Liu, X. 2025, Astronomy \& Astrophysics, 693, A168

\bibitem[{Liu \& Liu(2024)}]{liu2024unraveling}
Liu, B. \& Liu, X. 2024, Astronomy \& Astrophysics, 683, A51

\bibitem[{Liu {et~al.}(2016)Liu, Sachse, Spahn, \& Schmidt}]{liu2016dynamics}
Liu, X., Sachse, M., Spahn, F., \& Schmidt, J. 2016, Journal of Geophysical Research: Planets, 121, 1141

\bibitem[{Meech \& Jewitt(1988)}]{meech1988observations}
Meech, K. \& Jewitt, D. 1988, in Exploration of Halley’s Comet, Springer, 585--593

\bibitem[{Moreno(2025)}]{moreno2025cometary}
Moreno, F. 2025, Astronomy \& Astrophysics, 695, A263

\bibitem[{Moreno {et~al.}(2025)Moreno, Goetz, Aceituno, Casanova, Sota, \& Santos-Sanz}]{moreno2025dust}
Moreno, F., Goetz, C., Aceituno, F.~J., {et~al.} 2025, Monthly Notices of the Royal Astronomical Society, staf552

\bibitem[{Nicolini {et~al.}(2003)Nicolini, Cavicchio, \& Facchini}]{nicolini2003astroart}
Nicolini, M., Cavicchio, F., \& Facchini, M. 2003, Astroart 5.0, MSB Software

\bibitem[{Okuzumi {et~al.}(2012)Okuzumi, Tanaka, Kobayashi, \& Wada}]{okuzumi2012rapid}
Okuzumi, S., Tanaka, H., Kobayashi, H., \& Wada, K. 2012, The Astrophysical Journal, 752, 106

\bibitem[{Sekanina(2019)}]{sekanina20191i}
Sekanina, Z. 2019, arXiv preprint arXiv:1903.06300

\bibitem[{Sekanina(2024)}]{sekanina2024inevitable}
Sekanina, Z. 2024, arXiv preprint arXiv:2407.06166

\bibitem[{Stoer {et~al.}(1980)Stoer, Bulirsch, Bartels, Gautschi, \& Witzgall}]{stoer1980introduction}
Stoer, J., Bulirsch, R., Bartels, R., Gautschi, W., \& Witzgall, C. 1980, Introduction to numerical analysis, Vol. 1993 (Springer)

\bibitem[{Tang {et~al.}(2024)Tang, Wang, Lin, Yang, Zhang, Jia, \& Wang}]{tang2024spectrum}
Tang, Y., Wang, S., Lin, Z., {et~al.} 2024, Research Notes of the AAS, 8, 269

\bibitem[{Willmer(2018)}]{willmer2018absolute}
Willmer, C.~N. 2018, The Astrophysical Journal Supplement Series, 236, 47

\bibitem[{Ye {et~al.}(2023)Ye, Zhao, Li, Lu, Zhaori, Xu, Denneau, Erasmus, Fitzsimmons, Lawrence, {et~al.}}]{ye2023comet}
Ye, Q.-Z., Zhao, H., Li, B., {et~al.} 2023, Minor Planet Electronic Circulars, 2023

\end{thebibliography}
\bibliographystyle{aa}

\end{document}